\documentclass[iop]{emulateapj}
\usepackage{graphicx}
\usepackage{hyperref} 
\usepackage{amsmath}

\newcommand{\RA}[3]{\mbox{R.A.}={#1}$^{{\rm h}}${#2}$^{{\rm m}}${#3}$^{{\rm s}}$}
\newcommand{\decl}[3]{\mbox{Dec.}={#1}$^{\circ}${#2}\arcmin{#3}\arcsec}

\newcommand{\swift}{\emph{Swift}}
\newcommand{\fermi}{\emph{Fermi}}

\newcommand{\cm}[1]{~cm$^{#1}$}

\newcommand{\e}[1]{10$^{#1}$}
\newcommand{\ee}[1]{$\times$10$^{#1}$}

\newcommand{\fluence}{~erg\,cm$^{-2}$}

\newcommand{\nh}{N$_{\rm H}$}

\shorttitle{LAT detection of GRB~100728A}
\shortauthors{The Fermi Collaboration}

\begin{document}

\title{Detection of High-Energy Gamma-Ray Emission 
during the X-ray flaring activity in GRB~100728A}

\author{
A.~A.~Abdo\altaffilmark{2}, 
M.~Ackermann\altaffilmark{3}, 
M.~Ajello\altaffilmark{3}, 
L.~Baldini\altaffilmark{4}, 
J.~Ballet\altaffilmark{5}, 
G.~Barbiellini\altaffilmark{6,7}, 
M.~G.~Baring\altaffilmark{8}, 
D.~Bastieri\altaffilmark{9,10}, 
K.~Bechtol\altaffilmark{3}, 
R.~Bellazzini\altaffilmark{4}
% ...
}
% ...
\altaffiltext{2}{Center for Earth Observing and Space Research, College of Science, George Mason University, Fairfax, VA 22030, resident at Naval Research Laboratory, Washington, DC 20375}
\altaffiltext{3}{W. W. Hansen Experimental Physics Laboratory, Kavli Institute for Particle Astrophysics and Cosmology, Department of Physics and SLAC National Accelerator Laboratory, Stanford University, Stanford, CA 94305, USA}
\altaffiltext{4}{Istituto Nazionale di Fisica Nucleare, Sezione di Pisa, I-56127 Pisa, Italy}
\altaffiltext{5}{Laboratoire AIM, CEA-IRFU/CNRS/Universit\'e Paris Diderot, Service d'Astrophysique, CEA Saclay, 91191 Gif sur Yvette, France}
\altaffiltext{6}{Istituto Nazionale di Fisica Nucleare, Sezione di Trieste, I-34127 Trieste, Italy}
\altaffiltext{7}{Dipartimento di Fisica, Universit\`a di Trieste, I-34127 Trieste, Italy}
\altaffiltext{8}{Rice University, Department of Physics and Astronomy, MS-108, P. O. Box 1892, Houston, TX 77251}
\altaffiltext{9}{Istituto Nazionale di Fisica Nucleare, Sezione di Padova, I-35131 Padova, Italy}
\altaffiltext{10}{Dipartimento di Fisica ``G. Galilei", Universit\`a di Padova, I-35131 Padova, Italy}
% ...

\begin{abstract}
We present the simultaneous {\it Swift} and {\it Fermi} observations of the bright GRB~100728A 
and its afterglow. The early X-ray emission is dominated by a vigorous flaring activity 
continuing until 1\,ks after the burst. In the same time interval high energy emission is significantly detected by the {\it Fermi}/LAT. Marginal evidence of 
GeV emission is observed up to later times.
We discuss the broadband properties of this burst within both the internal and external shock scenarios,
with a particular emphasis on the relation between X-ray flares,
the GeV emission and a continued long-duration central engine activity
as their power source. 

\end{abstract}


\section{Introduction}\label{sec:intro}

The \fermi~Gamma-Ray Space Telescope, launched in June 2008,  
has taken the study of GRBs into an energy realm 
that so far has been poorly explored.
\fermi/LAT \citep[Large Area Telescope;][]{atwood09} observations of GRBs
allow for the first time a detailed study of the temporal and spectral behavior
at high energies ($>$100 MeV).
One of the most interesting features is the detection of a delayed and 
rapidly decaying high-energy emission, lasting hundreds to thousands of 
seconds longer than the observed sub-MeV $\gamma$-ray emission \citep{090902b,liv09,080916c}. 
Extended GeV emission, first hinted at in EGRET observations \citep{hurley94}, 
appears now as a common feature of \fermi/LAT bursts.
The nature of such long-lived high-energy emission is far from being established. 
One possibility is that it is generated via synchrotron radiation of the external 
forward shock \citep{kumar09,kumar10, ghisellini10}. An alternative scenario 
is that it reflects the gradual turn-off of the central
engine activity \citep{zhangbb10}. Such interpretations predict very different 
afterglow behaviors \citep{piran10, mimica10} and therefore can be directly verified 
through broadband (from optical/X-ray to GeV energies) early-time observations.   
To date, only one burst \citep[GRB~090510;][]{max10} 
of the 20 LAT detected GRBs has been simultaneously detected by the 
{\it Swift} multi-wavelength observatory \citep{gehrels04}.
In this case, an afterglow emission provides a likely
explanation of the broadband dataset \citep[e.g.][]{max10,corsi10}. 

In this Letter, we report on the \fermi/LAT detection of a temporally extended 
emission from GRB~100728A and the simultaneous {\it Swift} observations 
of an intense X-ray flaring activity. We further discuss the possibility 
that in the case of GRB~100728A the observed high-energy emission is related to X-ray flares and ultimately to the long-lasting activity of the inner engine. 
Observations and analysis are reported in \S~\ref{sec:data}; 
our results are discussed in \S~\ref{sec:res}; we draw
our conclusions in \S~\ref{sec:end}.    
Unless otherwise stated, the quoted errors are at the 90\% confidence level
and times refer to the \fermi/GBM trigger T$_0$. 
    
\section{Observations and Data analysis}\label{sec:data}

\subsection{Swift data}\label{sec:swift}

The bright GRB~100728A came into the \swift~field of view during a slew 
to a pre-planned target, when the trigger system is disabled. After the spacecraft
settled, the burst triggered the Burst Alert Telescope \citep[BAT;][]{bat05}  
on-board \swift~at 02:18:24 UT on 28th July 2010. \swift ~slewed immediately to the burst.
The two narrow field instruments, the X-ray Telescope \citep[XRT;][]{xrt05} and the Ultraviolet
Optical Telescope \citep[UVOT;][]{uvot05} began settled observations of the field $\sim$80\,s 
after the BAT trigger. A X-ray afterglow was promptly localized at a position
of \RA{05}{55}{2.01}, \decl{-15}{15}{19.1}  (J2000) with an uncertainty of 1.4$^{\prime\prime}$ \citep{gcnxrt}, 
while no counterpart was observed in the early UVOT unfiltered exposures 
down to a limiting magnitude $wh>$20.5 \citep[3 $\sigma$ confidence level;][]{gcnuvot}.

\swift~data were analyzed in a standard fashion; we refer the reader to 
\citet{evans07,evans10} for further details.
As shown in Fig.~1, the early X-ray afterglow (top panel) is characterized by
a series of bright X-ray flares superimposed on a power law decay ($\propto\,t^{-1.5}$).
Each flare can be described by a Fast-Rise Exponential Decay  (FRED) profile
(solid line) with $0.04<\Delta t/t < 0.2$.
In the same time interval, the BAT light curve (bottom panel) shows a long-lasting 
emission extending up to $\sim$800~s, with several peaks visible in coincidence with the
X-ray flares (vertical dot-dashed lines).

The postflare X-ray afterglow decays as a power law with slope $\alpha_2$=1.07$\pm$0.05,
which steepens to $\alpha_3$=1.63$\pm$0.07 at t$\sim$10 ks.
No significant spectral evolution is observed. 
The time-averaged photon index is $\Gamma=-2.07\pm0.09$.

By combining the simultaneous BAT and XRT observations, 
we performed a joint spectral analysis of the X-ray flares.
We modeled the absorption with two different components:
the former was fixed at the Galactic value of \nh=\e{21}\,\cm{-2} \citep{kalberla05};
the latter, representing the absorption local to the burst, 
was fixed to the value of \nh=2.6\ee{21}\,\cm{-2} derived from the late-time
(\e{4}-\e{6}\,s) afterglow spectrum.
This constraint prevents artificial \nh~variations 
caused by the intrinsic spectral evolution,
commonly observed in the brightest X-ray flares \citep{butler07}.
A strong spectral evolution is observed during the first 100\,s, 
showing a peak energy that softens from 95$\pm$15\,keV during the first flare 
(from T$_0$+167\,s to T$_0$+192\,s ) to less than 10\,keV in the following flares. Excluding the first harder episode, the time-averaged spectrum, from T$_0$+254\,s to T$_0$+854\,s, is well described by a Band function \citep{band93} ($\chi^2$=614 for 477 degrees of freedom, d.o.f.) with \mbox{$\alpha$=-1.06$\pm$0.11}, 
$\beta$=-2.24$\pm$0.02 and a peak energy E$_{\rm pk}$=1.0$^{+0.8}_{-0.4}$\,keV.


\begin{figure}[!t]
\centering
\includegraphics[angle=270,scale=0.35, viewport=1680 0 2230 700, clip]{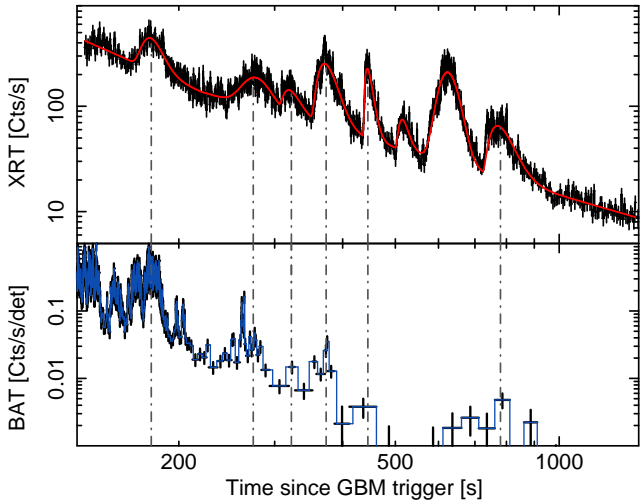}
\caption{{\it Top Panel:} Early XRT light curve of GRB~100728A. 
{\it Bottom Panel:} BAT mask-weighted light curve during the X-ray flaring activity. Several peaks are visible in correspondance of the
X-ray flares (vertical dot-dashed lines).}
\label{fig:lc}
\vspace{0.2cm}
\end{figure}


\subsection{Fermi data}\label{sec:fermi}
The \fermi/GBM triggered and located GRB~100728A at
02:17:31 UT, 53.6\,s before the \swift/BAT trigger (see \S~\ref{sec:swift}).
The GBM light curve shows a complex, multi-peaked structure 
with a duration $T_{90}\sim$163\,s in the 50-300 keV energy range
\citep{gcngbm}. 
A set of strong peaks is visible at $T_0$+170\,s, 
corresponding to the first flares detected by the XRT. 
No significant emission above the highly-variable background
level is detected on longer timescales. 

The time-averaged spectrum during the $T_{90}$ interval,  
from $T_0$+15\,s to $T_0$+178\,s
can be described with a Band function with the following
parameters: $\alpha$=-0.58$\pm$0.03, $\beta$=-2.73$^{+0.27}_{-0.18}$
and $E_{\rm pk}$=264$\pm$11\,keV (Castor C-statistics 865 for 351 d.o.f.). A power law function with an exponential
high-energy cutoff also provides an adequate description 
(C-statistics 885 for 352 d.o.f.). 
The event fluence (10-1000 keV) in the selected time interval is 
1.181$\pm$0.010\,\ee{-4}\,\fluence. The high fluence of this burst 
generated an Autonomous Repoint Request, which caused the
\fermi~satellite to slew to the GRB position.


 \begin{deluxetable*}{lcccc}
 \tablecolumns{5}
 \small
 \tablewidth{0pt}
 \tablecaption{LAT-Analysis Results} 
 \tablehead{ \colhead{}
 & \colhead{Time interval} & \colhead{Test Statistic}
& \colhead{Flux\tablenotemark{a}} & \colhead{$\Gamma_{\rm LAT}$\tablenotemark{b}}\\
\colhead{} & \colhead{(s)} &  \colhead{} & \colhead{(10$^{-6}$\,ph\,cm$^{-2}$\,s$^{-1}$)} & \colhead{}}
\hline
\startdata
 flare 1\dotfill& 167--192		&0&$<$28 &-- \\
 flare 2\dotfill& 254--304		&0&$<$12& \\
 flare 3\dotfill& 309--354		&11&$<$30&--\\
 flare 4\dotfill& 359--414		&0&$<$12 &--\\
 flare 5\dotfill& 439--474		&0&$<$15 &--\\
 flare 6\dotfill& 504--544			&11&$<$30&--\\
 flare 7\dotfill& 577--694			&0&$<$7&--\\
 flare 8\dotfill& 724--854			&0&$<$4&--\\
 \hline 
 \multicolumn{5}{c}{Time-integrated Search} \\
 \hline
 
pre-flares (prompt) \dotfill    & 0--167    & 5   & $<$18       & --    \\
post-flares\tablenotemark{c}\dotfill      	& 854--1654 & 10 & 0.7$\pm$0.5 & -1.4$\pm$0.4\\
X-ray flares \dotfill 		& 167--854  & 32  & 2.4$\pm$1   & -1.4$\pm$0.2\\
X-ray flares \dotfill 		& 254-854   & 27  & 2.0$\pm$1   & -1.3$\pm$0.3\\
flares 3 \& 6\dotfill    	& --	    & 22  & 9.6$\pm$5   & -1.2$\pm$0.3\\
flaring interval (excluding 3 and 6)  \dotfill 	&--	    & 17  & 1.6$\pm$1   & -1.6$\pm$0.4 
\enddata

\footnotetext{Fluxes in the 100 MeV - 50 GeV energy band.
The quoted errors are at the 68\% confidence level.\\ 
Upper limits are at the 95\% confidence level and were
calculated using the best fit photon index  \mbox{$\Gamma_{\rm LAT}$=-1.4}. 
}
\footnotetext{Here the photon index $\Gamma_{\rm LAT}$ is defined such that $dN/dAdEdt \propto E^{\Gamma_{\rm LAT}}$.}
\footnotetext{From diffuse-class LAT data. }

\label{tabLAT}
\end{deluxetable*}


\subsubsection{LAT Observations}

An unbinned likelihood analysis \citep{080825c} was used to search the LAT data for emission from GRB~100728A. As this study is part of a systematic search for high-energy (HE) emission from X-ray flares, 
a trials factor of 28 for the number of flares considered has to be taken into account in evaluating the detection significances.

Depending on the time window of interest the search was performed on transient-class data, optimally suited for short duration (tens of seconds) signal-limited studies, or diffuse-class data, best suited for detecting faint emission over longer timescales \citep{atwood09}.
The analysis included LAT events reconstructed within 15$^{\circ}$ around the XRT localization (\S~\ref{sec:swift}) with energies in the 100\,MeV--50\,GeV range. 
The GRB spectrum was modeled using a power law. No point source in the vicinity of the GRB (within 15$^\circ$) was bright enough to merit inclusion in the background model. 
The cosmic-ray background and the extragalactic gamma-ray background were estimated following the method described in \citet{080825c} for the transient-class searches and modeled as a single isotropic power law for the diffuse-class searches. The Galactic diffuse gamma-ray background was described by using the publicly available template produced by the LAT 
collaboration\footnote{ http://fermi.gsfc.nasa.gov/ssc/data/access/lat/BackgroundModels.html}. The background contribution from the Earth's albedo was negligible since the GRB position was far from the Earth's limb during all the time intervals analyzed.
Tests performed with different background models do not show any significant change in our results.

The results of our analysis are summarized in Table~\ref{tabLAT}. 
A time-resolved search performed on each individual flare did not find 
any significant excess in the LAT data, 
though a marginal evidence of emission (test statistic TS$>$10, single trial) 
is present during two of the flares (\#3 and \#6). 
A time-integrated search performed over the whole flaring interval led to a significant detection (TS=32 for transient-class events, 
TS=42 for diffuse-class events). 
In this time interval the total number of transient (diffuse) class events is 191 (29); according to the likelihood analysis the number of events associated with the GRB is $\sim$10 (6). 
The highest energy diffuse-class event detected during the flaring interval (at $T_0$+709\,s) and in spatial coincidence with the source has an energy of 1.68\,GeV. The probability of the LAT background producing an event with at least that energy and during the same interval is $\approx$7$\times$10$^{-4}$.
Events of higher energies, tens of GeV, are detected in the transient class dataset, but the high background rate does not allow us to significantly associate them with the GRB.
Our best localization of the LAT emission, derived from transient-class data analysis, is: 
\RA{05}{55}{49}, \decl{15}{03}{18}, with a statistical uncertainty of 0.1$^{\circ}$ (68\% confidence level)
and a systematic error of 0.2$^{\circ}$. This position is consistent with the \swift~localization (\S~\ref{sec:swift}).

In order to determine whether the LAT emission is temporally extended or mainly originated during the higher-significance
 flares (\#3 and \#6), we performed two stacked searches on the 
transient-class dataset: one aggregating the data during these two flares and one during the whole 
flaring period excluding the two flares. Emission at a comparable level and with consistent spectral properties is present during both time intervals (see Table~\ref{tabLAT}), therefore we conclude that the LAT emission extends over the whole flaring period. 
A cross-correlation analysis between the LAT (diffuse-class) and XRT lightcurves does not detect any significant temporal correlation or anti-correlation between the two datasets. Similar results
are obtained from the analysis of the transient-class events.

As shown in Table~\ref{tabLAT}, no emission is detected during the GRB prompt phase. The resulting upper limit is consistent with the extrapolation of the Band spectrum to the LAT energy range. 
Marginal evidence of emission (TS$\approx$10 for diffuse-class events) is present after the end of the observed X-ray flaring activity.

\section{Discussion}\label{sec:res}
Below we summarize the results that are relevant to address  the
origin of the GeV emission.

\begin{figure}[!b]
\centering
\vspace{0.5cm}
\includegraphics[angle=270,scale=0.36]{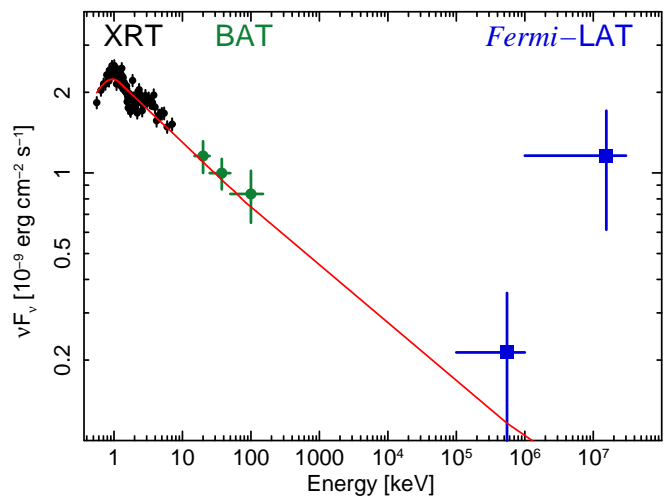}
\caption{ Spectral energy distribution of the X-ray flares 
including data from {\em Swift}/BAT, XRT and {\em Fermi}/LAT.
Error bars are at 1\,$\sigma$ confidence level.
The solid line shows the best fit model of
the joint XRT/BAT spectral fit  described in \S~\ref{sec:swift} and 
extrapolated to the {\em Fermi}/LAT energy range.
 As discussed in the text, the {\it Fermi}/LAT detection
is consistent within 3\,$\sigma$ with the extrapolation of the model.}
\label{fig:pha}
\end{figure}

\begin{itemize}

\item Significant GeV emission is found in the same interval where
the X-ray flaring activity is enhanced. However, the backgrounds and
limited statistics in the LAT data do not allow us to search for 
a one-to-one correlation
between the GeV emission and the single flare episodes. There is
marginal evidence of GeV emission after
the end of the flaring period.

\item The GeV flux is consistent with the extrapolation of the
power law describing the flare spectrum above 1 keV.
Assuming that an afterglow component is present below the flares
and that it has the same spectrum observed at later times, it is
found that the GeV flux is also consistent with the extrapolation
of this putative component. The  LAT  data exhibit a
harder spectrum than observed in X-rays (see Fig.~\ref{fig:pha}), 
though marginally consistent (within 3\,$\sigma$) 
with the X-ray spectral slope. 
\end{itemize}


The last result suggests that the HE emission can simply represent
the high energy tail of the synchrotron component. The presence of
an additional Inverse Compton (IC) component dominating over the 
synchrotron just above $\sim$1~GeV cannot be excluded, 
and it would be consistent with the observed flatter GeV spectrum. 
These deductions apply to whichever is the source of electron 
acceleration, internal or external shocks.

The discovery of HE emission in a time frame of vigorous flaring activity 
in X-rays lead us to consider firstly the association of the GeV emission 
with X-ray flares. We now discuss this scenario. 
Given the large number of flares, we
can exclude that a delayed external shock  is the dominant process
originating the X-ray flares \citep{gp07}. In fact, in this
model only a single outstanding flare, corresponding to the onset
of the afterglow from a long duration central engine, is produced.
We thus consider internal shocks from a long-lasting relativistic
outflow as the source of flares \citep[e.g.][]{zhang06}. 

In order to allow the GeV emission to be observed we require two conditions. First, the source has to be optically thin for pair production. 
By computing the optical depth for {e$^+$e$^-$} production \citep{lithwick01} by photons
of energy $E_{GeV}$ in GeV on the X-ray to GeV power law component
observed at about 300\,s we derive a lower limit on the Lorentz
factor:
\begin{equation}\label{gammagg}
\Gamma\ge\Gamma_{\gamma\gamma}\approx 30\,E_{GeV}^{1/6}\,t_{v}^{-1/6}
D_{28}^{1/3} \left(\frac{1+z}{2}\right)^{1/3},
\end{equation}\label{}
where we specialized the equation for a photon index 2.
The tightest constraints on $\Gamma$ are derived from the shortest time
scale for variability $t_v$ that can be associated to the
relativistic flow produced by the central source. 
 In the scenario of late internal shock X-ray flares and the 
GRB prompt emission are both related to the central engine activity.
We therefore consider that a variability timescale of ~ms typical of the 
prompt phase is a reasonable
possibility. In this case a single flare is produced by the superposition
of several internal shock events from a relativistic wind active for 
the whole duration of the flare.
By assuming $t_v=10^{-3}$\,s, a time scale similar to that characterizing the
prompt phase, one derives $\Gamma_{\gamma\gamma} \sim 115\,E_{GeV}^{1/6}$ for a 
typical redshift $z=1$. On the contrary,
if the flare is associated to a single internal shock event, i.e.
the interaction of two shells, then $t_v$ should be equal to the
flare duration, i.e. about 100\,s. In this case
$\Gamma_{\gamma\gamma} \sim 20\,E_{GeV}^{1/6}$. 

The lower boundary on the Lorentz factor derived above for the
flaring phase encompasses the range of values typical of the
prompt phase \citep{lithwick01,liangew10}, 
consistent with the notion that a long duration
relativistic outflow with a Lorentz factor of the order of $\approx$100
is producing both the prompt emission and the flares. 
In principle the parameters describing
the relativistic shocks ({$\epsilon_e, \epsilon_B, L, \Gamma,
t_v$}) can all be time dependent, i.e. be different during the
prompt and late flaring phases. On the other hand, one wishes to
reduce the number of variable parameters (Occam's razor). It goes
beyond the scope of this paper to find the best self-consistent
internal shock model reproducing both the prompt and X-ray flaring
phases. We just note the following. The model should be able to
reproduce a  peak energy that shifts from the $\approx$100 keV region during
the prompt phase to the keV range observed during X-ray flares.
Recalling that the peak of the synchrotron spectrum is given by 
\citep[e.g.][]{zhame02}:
\begin{equation}\label{num}
\nu_m\propto \epsilon_e^{3/2} \epsilon_B^{1/2} L^{1/2} \Gamma^{-2}
t_v^{-1},
\end{equation}
it follows that the decrease of the luminosity $L$ from the prompt to
the flare phase already accounts for a decrease of the peak
energy by a factor of 20, with all the other parameters remaining
constant. The further reduction that is needed can be obtained e.g.
by the very reasonable assumption that the magnetic field weakens
at the larger radii where flares are produced or by
a smaller contrast in the Lorentz factor between colliding shells \citep{barraud05}.

The second condition is derived by requiring that the maximum
energy at which the electrons are accelerated is large enough to
produce photons of energy E via synchrotron radiation:
\begin{equation}\label{gamma2}
\Gamma > 60 \left(\frac{1+z}{2} \right) (1+Y) E_{GeV},
\end{equation}
where $Y$ is the Compton y parameter.
This equation gives a condition on $\Gamma$ comparable to that
derived from Eq.~\ref{gammagg}. In conclusion we find that both the prompt
emission and the later X-ray flares and HE emission  can be
explained by internal shocks produced by a long duration central
engine with a Lorentz factor of $\approx 100$ and decreasing
luminosity.

This simple internal shock model predicts an emission that is
co-spatial and simultaneous in the X-ray and GeV ranges. 
On the other hand, we find
a marginal evidence of delayed HE emission. This is naturally predicted 
when the X-ray photons, produced by internal shocks at smaller radii, 
are upscattered to GeV energies via IC by the electron population of 
the forward shock \citep{wang06,fan08}.

Finally, given the quality of the present dataset, we cannot exclude
the possibility that the GeV emission is actually related to
an afterglow underlying the X-ray flares.
This requires that the afterglow onset takes place before 200\,s.
Such condition is satisfied when the Lorentz factor of the
relativistic flow at the beginning of the deceleration phase is:
 
\begin{equation}\label{gammaext}
\Gamma_{ext} > \left\{ \begin{array}{lr} 
171 \left(\frac{1+z}{2}\right)^{3/8}  \left(\frac{E_{54}}{n}\right)^{1/8}  & \mbox{ISM}, \\
74 \left(\frac{1+z}{2}\right)^{1/4} \left(\frac{E_{54}}{A^{*}}\right)^{1/4} & \mbox{Wind},
\end{array} \right.
\end{equation}
where A$^{*}$ is the density scaling factor in units of 5\ee{11}\,g\,cm$^{-1}$. 
In other GRBs, the HE emission has been
indeed associated to the forward shock synchrotron emission \citep{ghirlanda10,kumar10}, 
though the shorter time scales observed require much larger values of $\Gamma_{ext}$ than those derived here. 
We further explore the external shock scenario in GRB~100728A 
by analyzing the late time X-ray behavior. The afterglow
spectral and temporal laws are bound to obey specific relations
(the so-called closure relations, e.g. \citealt{zhangme04}), 
that depend upon the
density profile of the external medium, the jet opening angle, and
the relative position of the typical frequencies of the
synchrotron spectrum with respect to the observed range. 
Within the simple external shock model, the closest solution 
envisages a jet with a rather narrow opening angle of
$\approx$1-2\,deg expanding in a medium with a wind-like density profile,
though the lack of multi-wavelength afterglow observations 
does not allow us to firmly characterize the circumburst environment. 
This scenario is consistent with the lack of spectral variations before and
after the break at 10~ks, albeit in a wind-like medium a jet 
transition is expected to take place on much longer time scales 
\citep{kumar00}.  
In this scenario the cooling frequency falls below the X-ray band, and 
the early GeV emission (if afterglow) likely belongs to the same synchrotron regime. In this case, the LAT emission should display a similar decay slope
of $\sim$1.07 and a photon index of $\sim$-2.07, 
softer than the observed value of -1.4$\pm$0.2 (1\,$\sigma$) but still consistent
within the large uncertainty.

\section{Conclusion}\label{sec:end}

GRB~100728A is the second case to date with
simultaneous {\it Swift} and {\it Fermi} observations. 
High-energy gamma-rays are detected by the {\it Fermi}/LAT until 850\,s (TS=42)
and possibly continuing until 1600\,s (TS$\approx$10).
Very interestingly the early X-ray afterglow exhibits an 
intense and long-lasting flaring activity, visible both 
in BAT and XRT. 
Although an afterglow origin of the GeV emission cannot be 
excluded,  the presence of bright X-ray flares unveiled by 
{\it Swift} observations opens the possibility that a prolonged
central engine activity is powering the temporally extended HE emission
observed in this burst. 

Within the internal shock scenario a relativistic outflow
with a Lorentz factor of $\approx 100$ and decreasing
luminosity can explain the prompt emission, the later X-ray flares 
and HE emission. The presence of a delayed HE emission naturally
arises from IC scattering of low-energy flare photons off 
the relativistic electrons at the external forward shock radius.


\acknowledgements{}

We thank the referee for a careful reading of the paper. 
ET was supported by an appointment to the NASA
Postdoctoral Program at the Goddard Space Flight Center, 
administered by Oak Ridge Associated Universities 
through a contract with NASA.
The \fermi~LAT Collaboration acknowledges support from a number of agencies 
and institutes for both development and the operation of the LAT as well as 
scientific data analysis. These include NASA and DOE in the United States, 
CEA/Irfu and IN2P3/CNRS in France, ASI and INFN in Italy, MEXT, KEK, and JAXA
in Japan, and the K. A. Wallenberg Foundation, the Swedish Research Council 
and the National Space Board in Sweden. 
Additional support from INAF in Italy and CNES in France for science analysis 
during the operations phase is also gratefully acknowledged.


\vspace{5cm}

\bibliographystyle{aa}

\end{document}